\documentclass[12pt]{article}
\usepackage{blindtext}
\usepackage{scrextend}
\usepackage{hyperref}
\usepackage{amsmath}
\usepackage{amsfonts}
\usepackage{amssymb}
\usepackage{amsthm}                                                                                                                                                                                                                                        
\usepackage{times}

\newcommand{\ket}[1]{| #1 \rangle}

\title{In Defense of a ``Single-World'' Interpretation of Quantum Mechanics}
\author{Jeffrey Bub\\ \small Philosophy Department\\\small Institute for Physical Science and Technology\\\small Joint Center for Quantum Information and Computer Science\\  \small University of Maryland, College Park, MD 20742, USA}

\date{}
\normalsize
\begin{document}

\maketitle

\begin{abstract}
In a recent result,  Frauchiger and Renner argue  that if quantum theory accurately describes complex systems like observers who perform measurements, then ``we are forced to give up the view that there is one single reality.'' Following a review of the Frauchiger-Renner argument, I argue  that quantum mechanics should be understood \emph{probabilistically}, as a new sort of non-Boolean probability theory, rather than \emph{representationally}, as a theory about the elementary constituents of the physical world and how these elements evolve dynamically over time. I show that this way of understanding quantum mechanics is not in conflict with a consistent ``single-world'' interpretation of the theory.
\end{abstract}

\newpage
\section{Introduction}

In a recent ``no go'' result,  Frauchiger and Renner,\cite{FrauchigerRenner} argue that no ``single-world'' interpretation of quantum mechanics can be self-consistent, where a single-world interpretation is any interpretation that asserts, for a measurement with multiple possible outcomes, that just one outcome  actually occurs. The argument is a novel re-formulation of the ``Wigner's friend'' argument,\cite{Wigner} with a twist that exploits Hardy's paradox.\cite{Hardy} Frauchiger and Renner conclude that if quantum theory accurately describes complex systems like observers who perform measurements, then ``we are forced to give up the view that there is one single reality.''\cite{FrauchigerRenner2}

Following a review of the Frauchiger-Renner argument in \S{2}, I argue in \S{3} that quantum mechanics should be understood \emph{probabilistically}, as a new sort of non-Boolean probability theory, rather than \emph{representationally}, as a theory about the elementary constituents of the physical world, standardly particles and fields of a certain sort, and how these elements evolve dynamically as they interact over time.\cite{Wallace} In \S{4}, I show that this way of understanding quantum mechanics is not in conflict with a consistent ``single-world'' interpretation of the theory.

\section{The Frauchiger-Renner Argument}

Here's how the Frauchiger-Renner argument goes:  Alice measures an observable $A$ with eigenstates $\ket{h}_{A}, \ket{t}_{A}$ on a system in the state $\frac{1}{\sqrt{3}}\ket{h}_{A} + \frac{\sqrt{2}}{\sqrt{3}}\ket{t}_{A}$. One could say that Alice ``tosses a biased quantum coin'' with probabilities 1/3 for heads and 2/3 for tails. She prepares a qubit in the state $\ket{0}_{B}$ if the outcome is $h$, or in the state  $\frac{1}{\sqrt{2}}(\ket{0}_{B} + \ket{1}_{B})$ if the outcome is $t$, and sends it to Bob. When Bob receives the qubit, he measures a qubit observable $B$  with eigenstates $\ket{0}_{B}, \ket{1}_{B}$. After Alice and Bob obtain definite outcomes for their measurements, the quantum state of the combined quantum coin and qubit system is $\ket{h}_{A}\ket{0}_{B}$ or $\ket{t}_{A}\ket{0}_{B}$ or  $\ket{t}_{A}\ket{1}_{B}$, with equal probability.  At least, that's the state of quantum coin and qubit system from the perspective of Alice and Bob.

Now, the quantum coin and the qubit, as well as Alice and Bob, their measuring instruments and all the systems in their laboratories that become entangled with the measuring instruments in registering and recording the outcomes of the quantum coin toss and the qubit measurement, including the entangled environments, are just two big many-body quantum systems $S_{A}$ and $S_{B}$, which are assumed to be completely isolated from each other after Bob receives Alice's qubit. Consider two super-observers, Wigner and Friend, with vast technological abilities, who contemplate measuring a super-observable $X$ of $S_{A}$ with eigenstates $\ket{\mbox{fail}}_{A} =  \frac{1}{\sqrt{2}}(\ket{h}_{A} + \ket{t}_{A}), \ket{\mbox{ok}}_{A} = \frac{1}{\sqrt{2}}(\ket{h}_{A} - \ket{t}_{A})$, and a super-observable  $Y$ of $S_{B}$ with eigenstates $\ket{\mbox{fail}}_{B} =  \frac{1}{\sqrt{2}}(\ket{0}_{B} + \ket{1}_{B}),\ket{\mbox{ok}}_{B} =  \frac{1}{\sqrt{2}}(\ket{0}_{B} - \ket{1}_{B})$. 

To avoid unnecessarily complicating the notation by introducing new symbols for super-observables corresponding to observables, I'll use the same symbols $A$ and $B$ to represent super-observables of the composite systems $S_{A}$ and $S_{B}$ that  end up with definite values corresponding to the outcomes of Alice's and Bob's measurements on the quantum coin and the qubit, and I'll denote eigenstates of the super-observables $A$ and $B$ by the same symbols $\ket{h}_{A},\ket{t}_{A}$ and $\ket{0}_{B}, \ket{1}_{B}$ I used to represent eigenstates of the observables measured by Alice and Bob, with $\{h,t\}$ and $\{0,1\}$ representing the corresponding eigenvalues. (The alternative would be to use primes or some other notational device to denote super-observables and their eigenstates and eigenvalues, but this seems unnecessary, especially since the following argument concerns only super-observables. If the reader finds this confusing, simply add primes to all symbols from this point on.)

 Of course, such a measurement by the super-observers Wigner and Friend would be extraordinarily difficult to carry out in practice on the whole composite system, including Alice and Bob and their brain states, and all the systems in their environments, but nothing in quantum mechanics precludes this possibility. From the perspective of Wigner and Friend,  $S_{A}$ and $S_{B}$ are just composite many-body entangled quantum systems  that have evolved unitarily to a combined entangled state:
\begin{equation}
\ket{\psi}  =  \frac{1}{\sqrt{3}}(\ket{h}_{A}\ket{0}_{B} + \ket{t}_{A}\ket{0}_{B} +\ket{t}_{A}\ket{1}_{B}) \label{eqn:entangled}
\end{equation}
The outcomes of Alice's and Bob's measurements simply don't appear anywhere in the super-observers' description of events, so Wigner and Friend see no reason to conditionalize the state to one of the product states $\ket{h}_{A}\ket{0}_{B}$ or $\ket{t}_{A}\ket{0}_{B}$ or  $\ket{t}_{1}\ket{1}_{B}$. For Wigner and Friend, this would seem to require a suspension of unitary evolution in favor of an unexplained ``collapse'' of the quantum state.

But now we have a contradiction. The state $\ket{\psi}$ can also be expressed as:
\begin{eqnarray}
\ket{\psi} & = & \frac{1}{\sqrt{12}}\ket{\mbox{ok}}_{A}\ket{\mbox{ok}}_{B} - \frac{1}{\sqrt{12}}\ket{\mbox{ok}}_{A}\ket{\mbox{fail}}_{B} \nonumber\\
&& + \frac{1}{\sqrt{12}}\ket{\mbox{fail}}_{A}\ket{\mbox{ok}}_{B} + \sqrt{\frac{3}{4}}\ket{\mbox{fail}}_{A}\ket{\mbox{fail}}_{B}\\
& = & \sqrt{\frac{2}{3}}\ket{\mbox{fail}}_{A}\ket{0}_{B} + \frac{1}{\sqrt{3}}\ket{t}_{A}\ket{1}_{B}\\
& = & \frac{1}{\sqrt{3}}\ket{h}_{A}\ket{0}_{B} + \sqrt{\frac{2}{3}}\ket{t}_{A}\ket{\mbox{fail}}_{B}
\end{eqnarray}
From the first expression for $\ket{\psi}$, the probability is $1/12$ that Wigner and Friend find the pair of outcomes $\{\mbox{ok, ok}\}$ in a joint measurement of  $X$ and $Y$ on the two systems. But this outcome is \emph{inconsistent with any pair of outcomes for Alice's and Bob's measurements}. From the second expression, the pair $\{\mbox{ok}, 0\}$ has zero probability, so $\{\mbox{ok}, 1\}$ is the only possible pair of values for the super-observables $X, B$ if $X$ has the value $\mbox{ok}$. From the third expression, the pair $\{t, \mbox{ok}\}$ has zero probability, so $\{h, \mbox{ok}\}$ is the only possible pair of values for the super-observables $A, Y$ if $Y$ has the value $\mbox{ok}$. But the pair of values $\{h, 1\}$ for the super-observables $A$ and $B$ has zero probability in the state $\ket{\psi}$, so it does not correspond to a possible pair of measurement outcomes for Alice and Bob.

Both the observers, Alice and Bob, and the super-observers, Wigner and Friend,  apply quantum mechanics correctly. The argument depends only on (i) the one-world assumption, that a measurement has a single outcome, (ii)  the assumption that quantum mechanics applies to systems of any complexity, including observers, and (iii) self-consistency, in particular agreement between an observer and a super-observer. The surprising conclusion is that there is no consistent story that includes observers and super-observers: a pair of outcomes with finite probability, according to quantum mechanics, of the super-observers' measurements on the composite observer system is inconsistent with the observers obtaining definite (single) outcomes for their measurements.

The Alice-Bob measurements and the Wigner-Friend measurements could be separated by any time interval. As far as we know, there are no super-observers, but the actuality of a measurement outcome can't depend on whether or not a super-observer turns up at some point. It's the theoretical possibility of a super-observer that shows the inconsistency of the theory. One could put the problem this way: according to quantum mechanics, there can be no quantum measurements with definite (single) outcomes, because it is always possible that super-observers could turn up at some point, perhaps centuries after the Alice-Bob measurements when technology is sufficiently advanced to generate a Frauchiger-Renner contradiction.

\section{The Quantum Revolution}

Before considering the options in the light of the Frauchiger-Renner result, I want to review the genesis of quantum mechanics and argue that the theory should be understood \emph{probabilistically}, as a new sort of non-Boolean probability theory, rather than \emph{representationally}, as a theory about the elementary constituents of the physical world and their dynamical evolution.

Quantum mechanics began with Heisenberg's ``Umdeutung'' paper, \cite{Heisenberg} his proposed ``reinterpretation'' of physical quantities at the fundamental level as noncommutative. To say that the algebra of physical quantities is commutative is equivalent to saying that the idempotent elements form a Boolean algebra.  For the physical quantities or observables of a quantum system represented by self-adjoint Hilbert space operators, the idempotent elements are the projection operators, with eigenvalues 0 and 1. They represent yes-no observables, or properties (for example, the property that the energy of the system lies in a certain range of values), or propositions (the proposition asserting that the value of the energy lies in this range), with the two eigenvalues corresponding to the truth values, true and false. 

Heisenberg's insight amounts to the proposal that certain phenomena in our Boolean macro-world  that defy a classical physical explanation can be explained probabilistically as a manifestation of collective behavior at a non-Boolean micro\-level. The Boolean algebra of physical properties of classical mechanics is replaced by a family of ``intertwined'' Boolean algebras, one for each set of commuting observables, to use Gleason's term.\cite{Gleason} The intertwinement precludes the possibility of embedding the whole collection into one inclusive Boolean algebra, so you can't assign truth values consistently to the propositions about observable values in all these Boolean algebras. Putting it differently: there are Boolean algebras in the family of Boolean algebras of a quantum system, notably the Boolean algebras for position and momentum, or for spin components in different directions, that don't fit together into a single Boolean algebra, unlike the corresponding family for a classical system. 

The intertwinement of commuting and noncommuting observables in Hilbert space imposes objective pre-dynamic probabilistic constraints on correlations between events, analogous to the way in which Minkowski space-time imposes kinematic constraints on events. The probabilistic constraints encoded in the geometry of Hilbert space provide the framework for the physics of a \emph{genuinely indeterministic universe}. They characterize the way probabilities fit together in a world in which there are nonlocal probabilistic correlations that violate Bell's inequality up to the Tsirelson bound, and these correlations can only occur between intrinsically random events.\cite{Bub}  As von Neumann put it,\cite{vonNeumann2} quantum probabilities are ``sui generis.'' They don't quantify incomplete knowledge about an ontic state (the basic idea of ``hidden variables''), but reflect the irreducibly probabilistic relation between the non-Boolean microlevel and the Boolean macrolevel.

This means that quantum mechanics is quite unlike any theory we have dealt with before in the history of physics, and  there is no reason, apart from tradition, to assume that the theory can provide   the sort of representational explanation we are familiar with in a  theory that is commutative or  Boolean at the fundamental level. Quantum probabilities can't be understood in the Boolean sense as quantifying ignorance about the pre-measurement value of an observable, but cash out in terms of what you'll find if you ``measure,'' which involves considering the outcome, at the Boolean macrolevel, of manipulating a quantum system in a certain way. 

A quantum ``measurement'' is a bit of a misnomer and not really the same sort of thing as a measurement of a physical quantity of a classical system. It involves putting a microsystem, like a photon, in a situation, say a beamsplitter or an analyzing filter, where the photon is forced to make an intrinsically random transition recorded as one of  two  macroscopically distinct alternatives in a device like a photon detector. The registration of the measurement outcome at the Boolean macrolevel is crucial, because it is only with respect to a suitable structure of alternative possibilities that it makes sense to talk about an event as definitely occurring or not occurring, and this structure is a Boolean algebra.

From this perspective, Heisenberg's theory does not, without embellishment (e.g., as in Bohm's theory or the Everett interpretation) provide a representational story, but rather a way of deriving probabilities and probabilistic correlations with no causal explanation. They are ``uniquely given from the start'' as a feature of the non-Boolean structure, to quote von Neumann,\cite{vonNeumann1} related to the angles in Hilbert space, not measures over states as they are in a classical or Boolean theory.  

There is a rival way of thinking about quantum mechanics in terms of Schr\"{o}\-dinger's wave-mechanical version of the theory\cite{Schrodinger1} that lends itself to a representational interpretation. Here the notion of ``superposition'' appears as a new ontological category: propositions can true, false, and in the case of superpositions, indeterminate, or neither true nor false. The measurement problem then arises as the problem of explaining the transition from indeterminate to determinate as a dynamical evolution. But as Schr\"{o}dinger himself pointed out in a lecture to the Royal Institution in London in March, 1928, the wave associated with a quantum system evolves in an abstract, multi-dimensional representation space, not real  physical space, so  ``it is merely an adequate mathematical description of what happens'':\cite{Schrodinger2}
\begin{quote}
The statement that what \emph{really} happens is correctly described by describing a wave-motion does not necessarily mean exactly the same thing as: what \emph{really} exists is a wave-motion. We shall see later on that in generalizing to an \emph{arbitrary} mechanical system we are led to describe what really happens in such a system by a wave-motion in the generalized space of its co-ordinates ($q$-space). Though the latter has quite a definite physical meaning, it cannot very well be said to `exist'; hence a wave-motion in this space cannot be said to `exist' in the ordinary sense of the word either. It is merely an adequate mathematical description of what happens. It may be that also in the case of a single mass-point, with which we are now dealing, the wave-motion must not be taken to `exist' in \emph{too} literal a sense, although the configuration space happens to coincide with ordinary space in this particular simple case.
\end{quote}

The idea of a wave as a representation of quantum reality, and the associated measurement problem as the problem of accounting for the transition from indeterminate to determinate, ``from the limbo of potentialities to the clarity of actualities,''\cite{Ghirardi} as a dynamical process, continues to shape contemporary discussions of conceptual issues in the foundations of quantum mechanics. From the perspective adopted here, the later formalization of quantum mechanics by Dirac\cite{Dirac} in 1930 and von Neumann\cite{vonNeumann3} in 1932 as a theory of observables represented by operators on  a space of quantum states is fundamentally an elaboration of Heisenberg's ``Umdeutung'' rather than a wave theory. 

The really significant thing about a noncommutative mechanics  is the novel possibility of correlated events that are  \emph{intrinsically random}, not merely apparently random like coin tosses, where the probabilities of ``heads'' and ``tails'' represent an averaging over differences among individual coin tosses that we don't keep track of for practical reasons. This intrinsic randomness allows  \emph{new sorts of nonlocal probabilistic correlations} for ``entangled'' quantum states  of separated systems. Schr\"{o}dinger, who coined the term, referred to  entanglement (``Verschr\"{a}nkung'' in German) as ``\emph{the} characteristic trait of quantum mechanics, the one that enforces its entire departure from classical lines of thought.''\cite{Schrodinger3}  

The view that Hilbert space is fundamentally a theory of probabilistic correlations that are structurally different from correlations that arise in Boolean theories is, in effect, an information-theoretic interpretation of quantum mechanics. The classical theory of information was initially developed by Shannon to deal with certain problems in the communication of messages as electromagnetic signals along a channel such as a telephone wire. An information source produces messages composed of sequences of symbols from an alphabet, with certain probabilities for the different symbols. The fundamental question for Shannon was how to quantify the minimal physical resources required to represent messages produced by a source, so that they could be communicated via a channel and reconstructed by a receiver:\cite{Shannon}
\begin{quote}
The fundamental problem of communication is that of reproducing at one point either exactly or approximately a message selected at another point. Frequently the messages have meaning; that is they refer to or are correlated according to some system with certain physical or conceptual entities. These semantic aspects of communication are irrelevant to the engineering problem. The significant aspect is that the actual message is one selected from a set of possible messages. The system must be designed to operate for each possible selection, not just the one which will actually be chosen since this is unknown at the time of design.
\end{quote}

A theory of information in Shannon's sense is about the ``engineering problem'' of communicating messages over a channel efficiently. In this sense, the concept of information has nothing to with anyone's knowledge and everything to do with the stochastic or probabilistic process that generates the messages. What we have discovered in quantum phenomena is that the possibilities for representing, manipulating, and communicating information, encoded in the geometry of Hilbert space, are different than we thought, irrespective of what the information is about.

On this non-representational way of understanding quantum mechanics, as a non-classical theory of information or a new way of generating probabilities and probabilistic correlations between intrinsically random events, probabilities are defined with respect to \emph{a single Boolean frame}, the Boolean algebra generated by the ``pointer-readings'' of what Bohr referred to as the ``ultimate measuring instruments,'' which are ``kept outside the system subject to quantum mechanical treatment'':\cite{Bohr1}
\begin{quote}
In the system to which the quantum mechanical formalism is applied, it is of course possible to include any intermediate auxiliary agency employed in the measuring processes. \ldots The only significant point is that in each case some ultimate measuring instruments, like the scales and clocks which determine the frame of space-time coordination---on which, in the last resort, even the definition of momentum and energy quantities rest---must always be described entirely on classical lines, and consequently be kept outside the system subject to quantum mechanical treatment. 
\end{quote}

Bohr did not, of course, refer to Boolean algebras, but the concept is simply a precise way of codifying a significant aspect of what Bohr meant by a description ``on classical lines'' or ``in classical terms'' in his constant  insistence that  (his emphasis)\cite{Bohr2}
\begin{quote}
\emph{however far the phenomena transcend the scope of classical physical explanation, the account of all evidence must be expressed in classical terms.}
\end{quote}
 by which he meant ``unambiguous language with suitable application of the terminology of classical physics''---for the simple reason, as he put it, that we need to be able to ``tell others what we have done and what we have learned.'' Formally speaking, the significance of ``classical'' here as being able to ``tell others what we have done and what we have learned'' is that the events in question should fit together as a Boolean algebra. George Boole, who came up with the idea in the mid-1800's, introduced Boolean constraints on probability as ``conditions of possible experience.''\cite{Boole} 

It's not that unitarity  is suppressed at a certain level of complexity, where  non-Booleanity becomes Booleanity and quantum becomes classical. Rather, there is a macrolevel, which is Boolean, and there are actual events at the macrolevel. Any system, of any complexity, is fundamentally a quantum system and can be treated as such, in principle, which is to say that a unitary dynamical analysis can be applied to whatever level of precision you like. But at the end of the day, so to speak, some particular system, $M$, counts as the ``ultimate measuring instrument'' with respect to which an event corresponding to a definite measurement outcome occurs in an associated Boolean frame whose selection is not the outcome of a dynamical evolution described by the theory. The system $M$, or any part of $M$, can be treated quantum mechanically, but then some other system, $M^{\prime}$, treated as classical or commutative or Boolean, plays the role of the ultimate measuring instrument in any application of the theory.

The crucial assumption in this probabilistic interpretation of the theory is that  the outcome of a measurement is an intrinsically random event at the macro\-level, \emph{something that actually happens}, not described by the deterministic unitary dynamics, so outside the theory, or ``irrational'' as Pauli characterizes it (his emphasis):\cite{Pauli}
\begin{quote}
Observation thereby takes on the character of \emph{irrational, unique actuality} with unpredictable outcome. \ldots  Contrasted with this \emph{irrational aspect} of concrete phenomena which are determined in their \emph{actuality}, there stands the \emph{rational aspect} of an abstract ordering of the \emph{possibilities} of statements by means of the mathematical concept of probability and the $\psi$-function [I would say `by means of the geometry of Hilbert space'].
\end{quote}
Putting it differently, the  ``collapse,'' as a conditionalization of the quantum state, is something you put in by hand after recording the actual outcome. The physics doesn't give it to you.

\section{The Options}

What are the options in the light of the Frauchiger-Renner result? 

If the quantum state is interpreted representationally, as the analogue of the classical state in stipulating what's true and what's false, and we accept assumption (ii) of \S{2} and hence the universality of unitarity (so no ``collapse'' of the quantum state), the correct description of the composite system $S_{A} + S_{B}$ just before the super-observers' measurements is the entangled state (1), a superposition with several components, each associated with a different measurement outcome for Alice's and Bob's measurements. The entangled state is the source of the measurement problem, here presented as an inconsistency in the theory, given  the other assumptions in the Frauchiger-Renner argument. Dropping the ``one-world'' assumption (i) then leads to Everett's many-worlds interpretation. 

If we interpret the quantum state probabilistically, we seem to be forced to QBism, the quantum Bayesianism of Christopher Fuchs and Ruediger Schack.\cite{Fuchs1} The QBist rejects assumption (iii), the self-consistency assumption. On this view, all probabilities, including quantum probabilities, are understood in the subjective sense as the personal judgements of an agent, based on how the external world responds to  actions by the agent. For QBists, the Born rule ``is a normative statement \ldots  about the decision-making behavior any individual agent should strive for \ldots not a ``law of nature'' in the usual sense,'' and ``measurement outcomes \emph{just are} personal experiences for the agent gambling upon them.''\cite{Fuchs2} So there is no requirement that the perspective of an observer and a super-observer should be consistent.

There is another option, which is to reject assumption (ii)---not by restricting the universality of the unitary dynamics or any part of quantum mechanics, but by interpreting the quantum state probabilistically rather than representationally in the sense of \S3. Quantum probabilities  don't quantify incomplete knowledge about an ontic state, but reflect the irreducibly probabilistic relation between the non-Boolean microlevel and the Boolean macrolevel, expressed through the intrinsic randomness of events associated with the outcomes of quantum measurements.  On this option, what the Frauchiger-Renner argument shows is that  \emph{quantum mechanics, as it stands without embellishment, is self-contradictory if the quantum state is interpreted representationally.} The conclusion is avoided if we interpret the state probabilistically, with respect to a Boolean frame defined with respect to an ``ultimate measuring instrument'' or ``ultimate observer.''

 In a situation, as in the Frauchiger-Renner argument, where there are multiple candidate observers, there is a question as to whether Alice and Bob are ``ultimate observers,'' or whether only Wigner and Friend are  ``ultimate observers.'' The difference has to do with whether Alice and Bob perform measurements of the observables $A$ and $B$ with definite outcomes at the Boolean macrolevel, or whether they are manipulated by Wigner and Friend in unitary transformations that entangle Alice and Bob with systems in their laboratories, with no definite outcomes for the observables $A$ and $B$. What actually happens to Alice and Bob is different in the two situations.

If there are events at the macrolevel corresponding to definite measurement outcomes for Alice and Bob, then Alice and Bob  represent  ``ultimate observers'' and the final state of the combined quantum coin and qubit system is  $\ket{h}_{A}\ket{0}_{B}$ or $\ket{t}_{A}\ket{0}_{B}$ or $\ket{t}_{A}\ket{1}_{B}$, depending on the outcomes. If Wigner and Friend subsequently measure the super-observables $X, Y$ on the whole composite Alice-Bob system (so they are ``ultimate observers'' in the \emph{subsequent} scenario), the probability of obtaining the pair of outcomes $\{\mbox{ok}, \mbox{ok}\}$ is 1/4 for any of the product states  $\ket{h}_{A}\ket{0}_{B}$ or $\ket{t}_{A}\ket{0}_{B}$ or $\ket{t}_{A}\ket{1}_{B}$. After the measurement, the super-observables $A, B$ are indefinite, and so are the corresponding quantum coin and qubit observables. There is no contradiction  because the argument from the entangled state no longer applies. If Wigner and Friend are ``ultimate observers'' but not Alice and Bob, there are no events at the macrolevel corresponding to definite measurement outcomes for Alice and Bob and the state is the entangled state (1). The probability of Wigner and Friend finding the pair of outcomes $\{\mbox{ok}, \mbox{ok}\}$ is 1/12, but there is no contradiction because there are no measurement outcomes for Alice and Bob.

There is a factual difference at the Boolean macrolevel between the two cases.  If Alice and Bob are ``ultimate observers'' in the application of quantum mechanics to the scenario, the probability of the super-observers finding the pair of outcomes $\{\mbox{ok}, \mbox{ok}\}$ is 1/4. If Wigner and Friend are ``ultimate observers'' but not Alice and Bob,  then the probability of Wigner and Friend finding the pair of outcomes $\{\mbox{ok}, \mbox{ok}\}$ is 1/12. The difference between the two cases--- the difference between probability 1/4 and probability 1/12---is an objective fact at the macrolevel.

Special relativity, as a theory about the structure of space-time, provides an explanation for length contraction and time dilation through the geometry of Min\-kow\-ski space-time, but that's as far as it goes.  This explanation didn't satisfy Lorentz, who wanted a dynamical explanation in terms of forces acting on physical systems used as rods and clocks.\cite{Janssen} Quantum mechanics, as a theory about randomness and nonlocality, provides an explanation for probabilistic constraints on events through the geometry of Hilbert space, but that's as far as it goes. This explanation doesn't satisfy Bohmians or Everettians, who insist on a representational story about how nature pulls off the trick of producing intrinsically random events at the macrolevel, with nonlocal probabilistic correlations constrained by the Tsirelson bound.

Such a representational story comes at a price that shapes the future direction of physics. Do we really want to give up the concept of measurement as a procedure that provides information about the actual value of an observable of a system to preserve the ideal of representational explanation in physics, as the Everett interpretation proposes? It seems far more rational to accept that if current physical theory has it right, the nature of reality, the way things are, limits the sort of explanation that a physical theory provides.

\section*{Acknowledgements}
Although we probably don't agree, thanks to Matt Leifer for clarification of the Frauchiger-Renner argument, and to Michel Janssen and Michael Cuffaro for some really helpful input on early drafts of this paper.

\end{document}